# Combinatorial search of superconductivity in Fe-B composition spreads

Kui Jin[1,*], Richard Suchoski[2], Sean Fackler[2], Yi Zhang[3], Xiaoqing Pan[3], Richard L. Greene[1], and Ichiro Takeuchi [2]

[1]CNAM and Department of Physics, University of Maryland, College Park, Maryland 20742, USA
[2]Department of Materials Science and Engineering, University of Maryland, College Park, Maryland 20742, USA
[3]Department of Materials Science and Engineering, University of Michigan, Anne Arbor, Michigan, 48109, USA

*Current address: Institute of Physics, Chinese Academy of Sciences, Beijing 100190, China

Abstract: We have fabricated Fe-B thin film composition spreads in search of possible superconducting phases following a theoretical prediction by Kolmogorov *et* al.[1] Co-sputtering was used to deposit spreads covering a large compositional region of the Fe-B binary phase diagram. A trace of superconducting phase was found in the nanocrystalline part of the spread, where the film undergoes a metal to insulator transition as a function of composition in a region with the average composition of $FeB_2$. The resistance drop occurs at 4K, and a diamagnetic signal has also been detected at the same temperature. The superconductivity is suppressible in the magnetic field up to 2 Tesla.

Throughout the one-hundred-year history of superconductivity, search and discovery of new superconductors has largely relied on serendipitous trial-and-error processes. Recently, there have been several computational predictions of new superconductors based on screening of known crystal structures in the inorganic crystal structure database (ICSD).[1-3] In one study, Kolomogorov *et al* [1] have carried out high-throughput screening by calculating formation enthalpies for over 40 borides and carbides and found that $FeB_4$ with an oP10 structure could be a phonon-mediated superconductor. The calculations showed that the strength of electron- phonon coupling ($\lambda$) was mostly generated by the mixed Fe-B modes, and the logarithmic average of the phonon frequencies $<\omega>$ was found to be close to that of $MgB_2$. These two quantities can be used to determine the superconducting transition temperature in the McMillan formula[4] and give an estimated $T_c$ of 15-20 K for $FeB_4$. Moreover, the authors also discuss the intriguing possibility of spin fluctuations playing a critical role in the pairing mechanism of this system.[5] Given the diversity in compositions and properties of the broad family of Fe-based superconductors uncovered to date,[6] it is important to investigate possibilities of other Fe containing superconductors.

High boron concentration compounds are in general notorious for their difficulty in proper synthesis owing to the high melting point of boron, and such compounds are often metastable. To get around this problem, one can synthesize bulk compounds under high pressure.[7] Another synthesis route is to make use of non-equilibrium thin film processes such as pulsed laser deposition and magnetron sputtering. In this study, we have utilized thin film co-sputtering composition spreads to search for possible superconducting phases in the Fe-B binary system. Previously, we have successfully implemented continuous composition spreads to discover new compositions of ferromagnetic shape memory alloys, morphotropic phase boundary piezoelectrics, giant magnetostrictive materials, etc.[8-12] Combinatorial libraries[13] and composition spreads have also been used in the past to demonstrate simultaneous synthesis of multiple superconducting compounds as well as to map compositional phase diagrams of known superconductors.[14-16]

Fe and B targets were co-sputtered to generate mapping of a large fraction of the binary compositional phase space across 3" Si wafers with a 200 nm $SiO_2$ layer on top. Fe and B were sputtered at 10 W (DC) and 100 W (RF), respectively. The composition spread films were deposited under Ar atmosphere of 15-70 mTorr with deposition temperatures between 200 to 650℃. Some of the spread wafers were annealed in-situ in vacuum afterward. The typical film thickness was 35-100 nm. Wavelength dispersive spectroscopy (WDS) mapping revealed that the average composition measured at different spots on the wafer ranged from almost pure Fe to $FeB_x$ with x > 60 over the 3" diameter area. The composition mapping information was then used to narrow the search for superconducting regions from approximately $FeB_2$ and $FeB_4$, and the wafers were diced into 1 $cm^2$ chips for simultaneous resistance versus temperature measurements of multiple composition spots.

Figure 1 shows the photograph of one wafer (deposited at 550℃) taken under ambient light. The arrows indicate the direction of compositional variation. The film on the Fe-rich side of the wafers shows good electrical conduction at room temperature with resistivity in the range of 10~50 μΩcm, and displays a metallic behavior as a function of temperature. This is consistent with the presence of the α-Fe phase over a large composition range as confirmed by x-ray diffraction mapping (not shown) of the spread. The (110) peak of α-Fe phase gets broader and broader with diminishing intensity toward the B-rich region of the spread and eventually disappears. This is also consistent with the fact that due to the extremely high crystallization temperature of B, the films here (annealed at up to 650℃) become more and more nanocrystalline-like with increasing average B concentration, and the films in the B-rich region are mostly amorphous.

The films on the B-rich side of the wafers display high resistivity at room temperature (100 ~1000 μΩcm), and the films are semiconducting to insulating at low temperatures, as discussed below. The clear color change seen in Fig. 1 between the Fe-rich metallic region and the B-rich insulating/semiconducting region takes place in the broad compositional transition between metallic films and semiconducting to insulating films. We note that the area of the wafer where the average composition ranges from $FeB_2$ to $FeB_4$ is in this transition region, where the films are also mostly nanocrystalline.

A number of diced 1 cm × 1 cm chips from this transition region were measured using the 64-pogo pin, 16-spot simultaneous resistance versus temperature measurement set up [Fig. 2(a)]. The entire area covered in Fig. 2(a) corresponds to the outlined boxed area of the wafer in Fig. 1. The pins are arranged in such a way [Fig. 2(b)], so that 16 multiplexed channels of temperature dependent resistance (R(T)) can be measured at 4 × 4 evenly spaced 1 mm × 1 mm areas in a Van der Pauw configuration. Figure 2(a) shows multiple R(T) curves, taken at 16 spots per chip from different selected chips cut out from one Fe-B composition spread wafer (annealed at 550℃), overlaid over the photograph of the wafer where the chips came from. A quick glance at all the curves across the wafer gives a clear trend: transition from the metallic Fe-rich crystalline region to the semiconducting/insulating B-rich nanocrystalline/amorphous region. One spot on one chip in the middle of the resistance transition region was found to display a sharp but partial resistance drop indicative of presence of superconductivity below 10 K. The same chip was then mounted on an AC susceptibility probe, and a clear but weak diamagnetic signal was observed starting at about 9 K [Fig. 2(c)]. The measured average composition of the spot which displayed the resistance drop was approximately $FeB_2$.

In order to further confirm the existence of superconductivity, more careful measurements were carried out on further separated smaller pieces (1-2 mm$^2$ area) of the same 1 cm$^2$ chip using a Quantum Design PPMS. As seen in Fig. 3(a), the resistance decreases abruptly once temperature decreases below 4 K, and then it stops decreasing at 2.5K in zero field. Clear diamagnetism was also observed as shown in Fig. 3(b). The partial resistance drop and the small diamagnetic signal indicate an incomplete superconducting transition. One reason for this might be that the superconducting phase is only present in a filamentary-like manner as a small volume fraction of the material. Such signature of possible presence of superconductivity was observed in chips from two composition spread wafers in the present study.

To obtain a robust evidence of superconductivity, we have studied the resistance behavior in magnetic fields applied in the out of the film plane direction [Fig. 3(c)]. As the magnetic field is increased, the drop in resistance is suppressed, a distinct feature for a superconductor. At the same time, negative magnetoresistance appears, caused either by localization[17] or by spin scattering.[18] The upper critical field ($B_{c2}$) is defined as the starting transition point of R(T) in fields, and plotted against the temperature in Fig. 3(d). We use a formula derived from the Ginsburg-Laudau theory, $B_{c2}(T)= B_{c2}(0)[1-(T/T_c)^2]/[1+(T/T_c)^2]$, to fit the $B_{c2}(T)$ data. The experimental data can be well described with a reasonable zero temperature upper critical field $B_{c2}(0)$ of 2 Tesla. This suggests that the superconductor we found is a conventional type II BCS superconductor.

To gain insight into the structural origin of the superconducting phase, we have carried out high-resolution cross-sectional transmission electron microscopy on the film spot which displayed superconductivity (Fig. 4). As seen in Fig. 4(a), the film consists largely of nanocrystalline grains and amorphous regions in between. Fig. 4(b) shows weak but discernible diffraction rings taken from this region. A magnified view inside one of the nanocrystalline grains [Fig. 4(c), left] indeed reveals crystalline features. The crystal structure of this grain was identified to be oS8. The boxed regions from Fig. 4(c) (left) are further magnified, and atomic images were obtained: the top right image and the bottom right image are along the <001> zone axis and along the <111> zone axis, respectively.

The oS8 is the known structure of FeB, but FeB is not known to be a superconductor. We do not rule out the possibility that the composition of the nanocrystalline phase seen here deviates from stoichiometric FeB. We note that the observed phase separation into nanocrystalline grains and the amorphous regions is due to the high crystallization temperature of B compounds and the natural quenching process from the hyperthermal plasma during sputtering. The complex nature of this process has likely resulted in local compositional heterogeneity, and the exact local compositional variation and distribution is difficult to determine. Indeed, the diffraction rings indicate that the lattice constants are very close to those of FeB, but there are slight discrepancies [Fig. 4(b)]. Therefore it is possible that the observed oS8 phase is responsible for the observed superconducting behavior.

Recently Gou *et al* have reported that $FeB_4$ in the oP10 ground state has been synthesized at pressures above 8 GPa and high temperature up to 2000 K[7]. The superconductivity at 3 K was verified by diamagnetism and the isotope effect, but the sample was too small for measuring the resistive transition. Given the small fraction of the volume which was found to exhibit superconducting behavior in our samples, there may also be a small fraction of superconducting oP10 $FeB_4$ in our film. Outside of the predicted $FeB_4$, there are no known superconducting phases in the Fe-B phase diagram, to the best of our knowledge. Even accounting for the remote possibility of diffusion taking place between the film and the substrate (which is not expected for the processing temperature used here), all known phases of superconductors consisting of Fe,

B, Si, O and Pd (electrode layer) have much lower transition temperatures and critical fields[19,20] than the ones observed here. We see no evidence of such diffusion from our microscopy study either. Further work is ongoing to identify and isolate the single superconducting phase in our thin film samples.

In summary, a composition spread technique was employed to search for superconductivity in the Fe-B binary phase. A partial superconducting transition was reproducibly found in the part of the spread where the spread film shows a metal to insulator transition, and where the average composition was approximately $FeB_2$. The resistive superconducting transition temperature was 4 K, and by fitting to a formula based on the Ginsburg-Landau theory, the upper critical field was found to be 2 Tesla.

This work was funded by AFOSR-MURI Grant FA9550-09-1-0603.

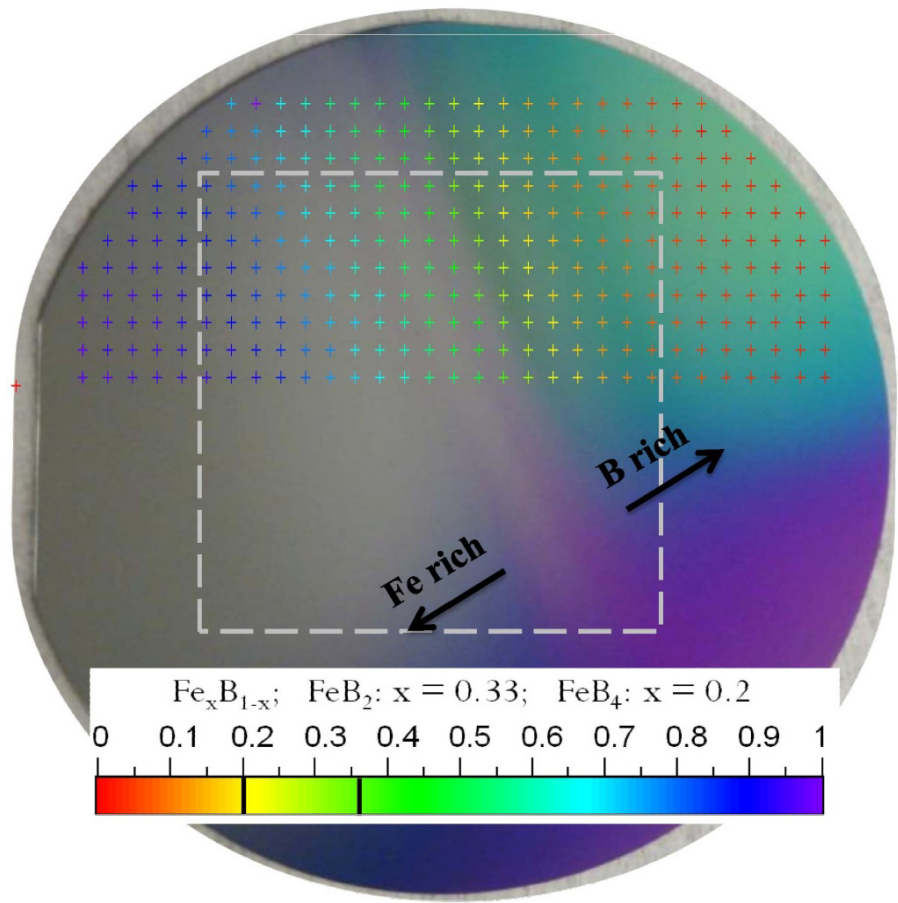

FIG.1. (Color online) A 3" Fe-B composition spread wafer. The spread film was synthesized by co-sputtering. Wavelength dispersive spectroscopy was used to map the compositions. The color of the cross symbols on the film shows a continuous average composition change from pure iron to $FeB_x$ (x > 60) as indicated by the color bar. The photograph of the spread film (taken under ambient light) displays a transition in color. The region of the color change in the middle of the wafer corresponds to area where the composition changes roughly from $FeB_2$ to $FeB_4$ (marked in the color bar with black lines). The boxed region outlined in white is the area covered in Fig. 2a.

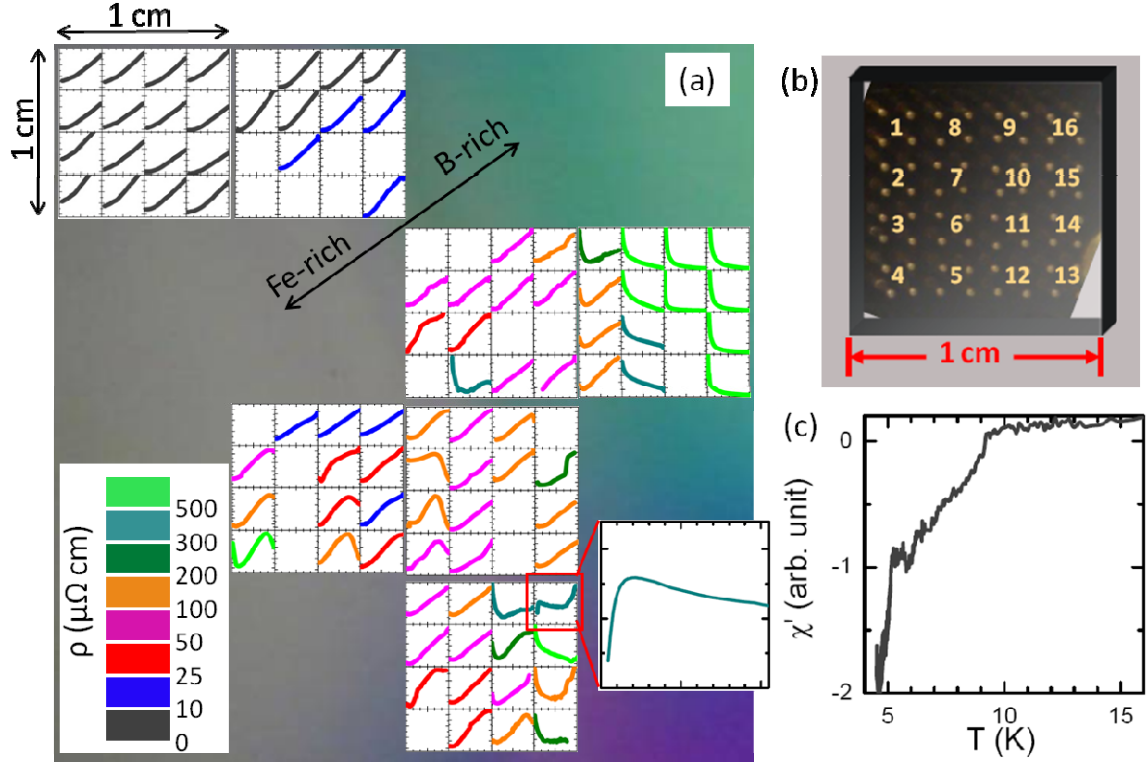

FIG.2. (Color online) Mapping of the temperature dependence of resistivity on the Fe-B composition spread film. The wafer was diced into 1cm$^2$ chips. On each chip ρ(T) of 16 channels can be measured simultaneously, and data are plotted at where the spots were located. For each plot, the vertical scale is resistivity (in μΩ cm). The resistivity is denoted by different colors as seen in the color bar. The horizontal scale is temperature from 0 to 300K. In going from Fe-rich (left lower ) to B-rich (upper right ) regions, the overall trend is that resistivity increases and shows a broad metal-to-insulator transition around the FeB$_2$ - FeB$_4$ region as seen in Fig. 1. One spot on one chip near the metal-to-insulator region (outlined in red) displayed a sharp resistivity drop indicative of presence of superconductivity below 10 K, which is zoomed in and replotted from 0 to 100 K. (b) A 64-pogo-pin-array probe, which is designed for 1 cm$^2$ chips. (c) A clear, but weak diamagnetic signal was also observed starting at about 9 K measured by AC susceptibility on the same chip..

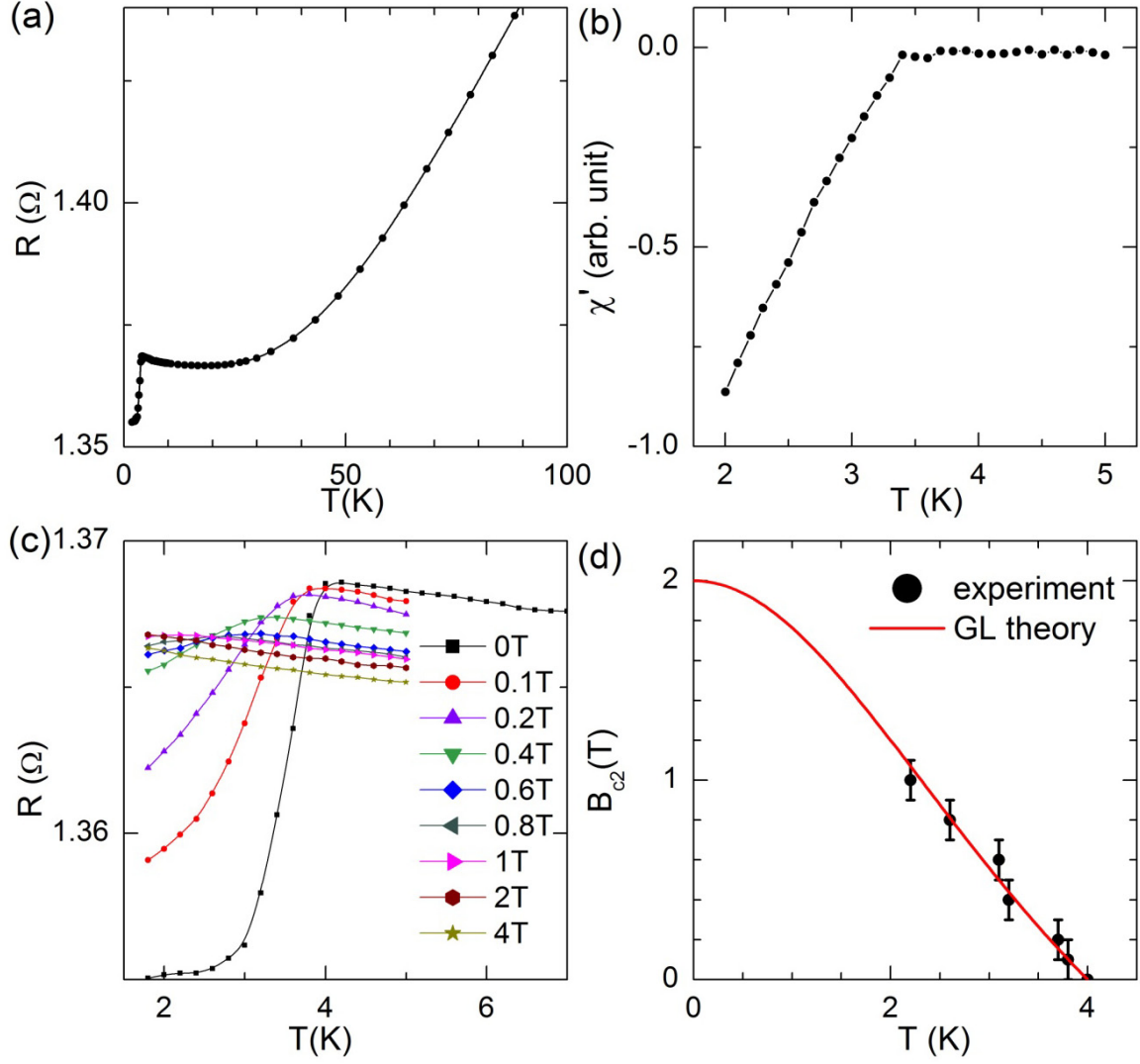

FIG.3. (Color online) (a) R(T) of one piece of Fe-B spread film cut from the same chip shown in Fig.2. (b) AC susceptibility measured on the same spot. (c) R(T) in different magnetic fields (perpendicular to the film plane). (d) Temperature dependence of upper critical field extracted from the R(T) data in (c), which obeys $B_{c2}(T)=B_{c2}(0)[1-(T/T_c)^2]/[1+(T/T_c)^2]$ and gives $B_{c2}(0) = 2$ Tesla.

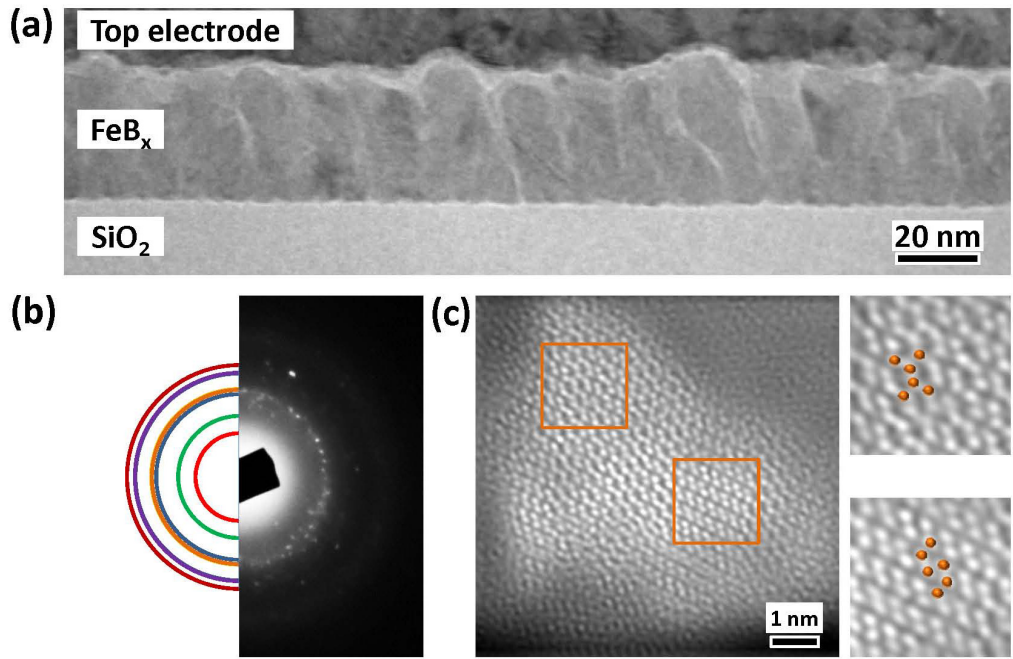

FIG.4. (Color online) (a)Cross-section TEM image of the Fe-B sample which shows superconductivity. The film consists largely of nanocrystalline grains and amorphous regions in between. The top electrode is a layer of Pd. (b) Selected-area electron diffraction pattern. Right half: experimental data; Left half: standard FeB-oS8 diffraction rings. (c) Left: The zoom-up inside one of the nanocrystalline grains, revealing crystalline features. The top right and the bottom right images are atomic images along the <001> and <111> zone axes, respectively.